\documentclass[conference]{IEEEtran}
\IEEEoverridecommandlockouts
\usepackage{cite}
\usepackage{amsmath,amssymb,amsfonts}
\usepackage{graphicx}
\usepackage{textcomp}
\usepackage{xcolor}
\usepackage{subfigure}
\usepackage{algorithm}
\usepackage{algorithmic}
\usepackage{mathrsfs}
\usepackage{subcaption}
\usepackage{multirow}
\usepackage{adjustbox}
\usepackage[labelformat=simple, labelsep=colon]{subcaption}

\def\BibTeX{{\rm B\kern-.05em{\sc i\kern-.025em b}\kern-.08em
    T\kern-.1667em\lower.7ex\hbox{E}\kern-.125emX}}
    
\begin{document}

\title{Analysis of Blockchain Assisted Energy Sharing Algorithms with Realistic Data Across Microgrids}


\author{
    Abdulrezzak Zekiye$^\star$, Ozan Sina Bankaoglu$^\star$, Ouns Bouachir$^\dag$, \"Oznur \"Ozkasap$^\star$, Moayad Aloqaily$^\ddag$ \vspace{5px} \\
    {$^\star${Ko\c{c} University, Department of Computer Engineering, İstanbul, Türkiye}}\\
    {$^\dag$College of Technological Innovation (CTI), Zayed University, UAE}\\
    {$^\ddag$Mohamed bin Zayed University of Artificial Intelligence (MBZUAI), UAE}\\
    Emails:{$^\star$\{azakieh22, obankaoglu18, oozkasap\}@ku.edu.tr, $^\dag$ouns.bouachir@zu.ac.ae, $^\ddag$moayad.aloqaily@mbzuai.ac.ae
    }
}

\IEEEoverridecommandlockouts

\maketitle
\IEEEpubidadjcol

\begin{abstract}
With escalating energy demands, innovative solutions have emerged to supply energy affordably and sustainably. Energy sharing has also been proposed as a solution, addressing affordability issues while reducing consumers' greed. In this paper, we analyse the feasibility of two energy sharing algorithms, centralized and peer-to-peer, within two scenarios, between microgrids within a county, and between microgrids across counties. In addition, we propose a new sharing algorithm named Selfish Sharing, where prosumers take advantage of consumers' batteries in return for letting them consume part of the shared energy. 
The results for sharing between microgrids across counties show that the dependency on the grid could be reduced by approximately 5.72\%, 6.12\%, and 5.93\% using the centralized, peer-to-peer and selfish sharing algorithms respectively, compared to trading only. The scenario of sharing between microgrids within a county has an average decrease in dependency on the grid by 5.66\%, 6.0\%, and 5.80\% using the centralized, peer-to-peer and selfish algorithms respectively, compared to trading without sharing. We found that trading with batteries and the proposed sharing algorithms prove to be beneficial in the sharing between microgrids case. More specifically, the case of trading and sharing energy between microgrids across counties outperforms sharing within a county, with P2P sharing appearing to be superior. 
\end{abstract}

\begin{IEEEkeywords}
Energy sharing, microgrids, blockchain, energy trading.
\end{IEEEkeywords}

\section{Introduction}
\begin{figure}
    \centering
    \includegraphics[width=1\linewidth]{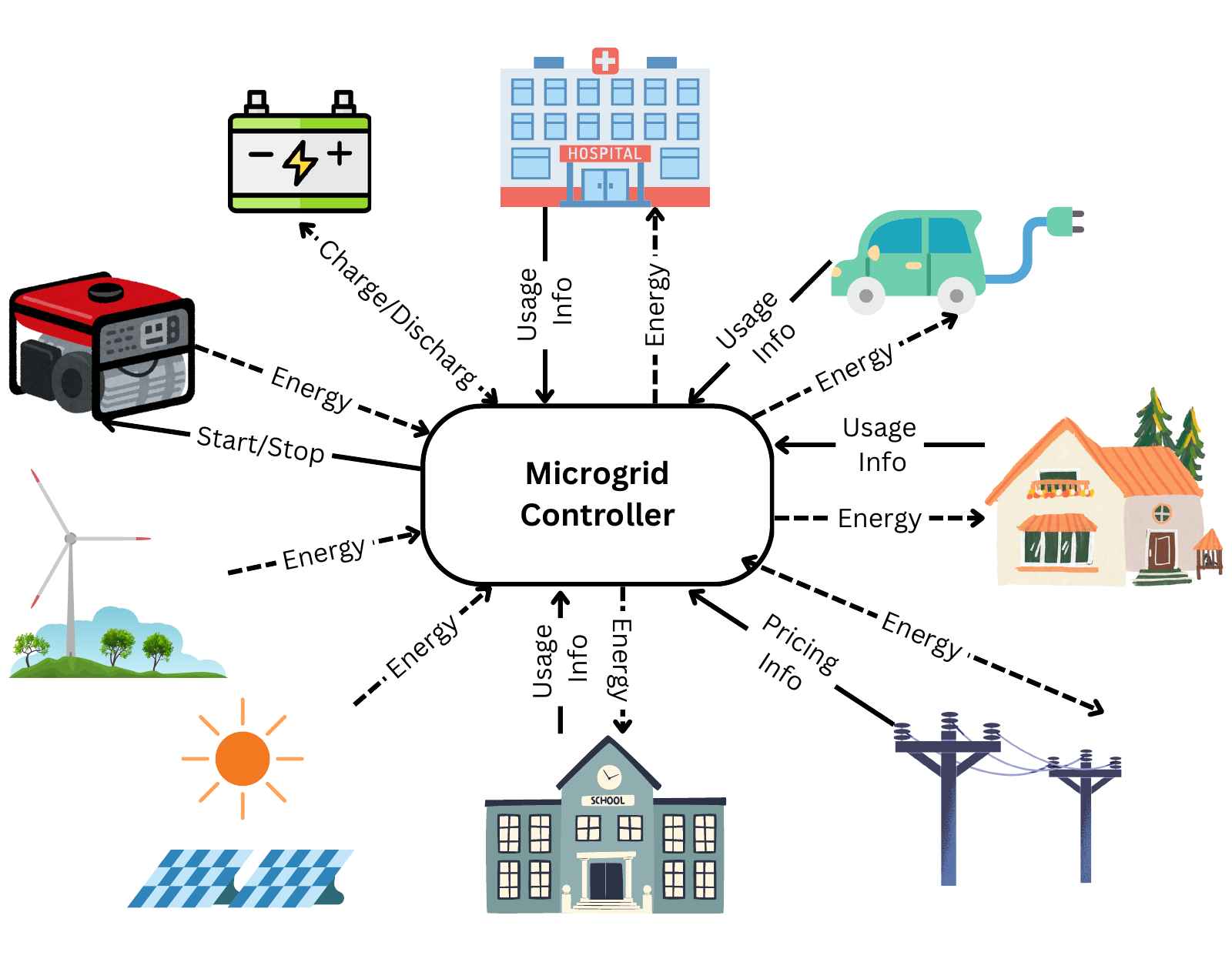}
    \caption{Microgrid coordinating between distributed energy sources, loads, and utility grid.}
    \label{fig:microgrid}
\end{figure}

The increasing energy need and the rising environmental issues that raising from utility grids, led more consumers to install energy generation tools, such as solar panels, to enable them to use energy in a more clean and affordable way. In Europe, the usage of renewable energy is targeted to be about 45\% in 2030 \cite{europa_renewable}. In another interesting study, the number of homes with solar panels in the United States is expected to be 16.8 millions in 2032, compared to only 3.6 millions in 2022 \cite{statista_solar_homes}. Those consumers might still need energy and might generate energy more than what they need. In the case of generating energy more than needed, the consumers become energy providers, or what is called prosumers. With the spread of prosumers that are considered as distributed energy sources, microgrid term appeared. A microgrid brings together distributed power sources, loads, energy storage devices, and control devices to create a unified and manageable power supply system \cite{zhou2015overview}. 
Figure \ref{fig:microgrid} illustrates how the microgrid controller coordinates between the distributed energy resources, loads, and utility grid. Microgrid controller decides when to sell or buy from the utility grid, charges/discharges the energy storage, utilizes the renewable and distributed energy resources efficiently, and sends energy to loads when needed. 

Peer-to-peer (P2P) energy trading between prosumers and consumers within or between microgrids was introduced to facilitate the transfer of energy between peers in more affordable way \cite{long2017peer}. Some solutions utilized blockchains and smart contracts to provide the trading service in a decentralized and more reliable manners such as \cite{perk2020joulin}, where blockchain could be considered as an immutable, decentralized, distributed, auditable and transparent ledger that is managed by a consensus protocol \cite{zheng2018blockchain}. Smart contracts on the other hand are programs that are hosted on the blockchain, where a group of people trusts the program to do some operations for them in a decentralized and transparent way \cite{zou2019smart}. 

Although P2P energy trading aims to offer the energy at a more affordable price, there are still some consumers or even prosumers not able to generate enough energy for their need, and not able to afford it as well. For instance, in Spain during the year 2021, energy poverty was 9.3\% of the households, meaning approximately tenth of the households consumed energy less than the national median \cite{statista_energy_poverty_spain}. Decreasing energy poverty and enabling consumers to access energy when they need it and not being able to afford it have been addressed in studies such as \cite{cali2021novel}, \cite{bouachir2022federatedgrids}, and \cite{zekiye2024blockchain}. In this paper, we examine the effectiveness of previously proposed battery-based sharing algorithms between microgrids in two different scenarios. In addition, we propose a new sharing algorithm, namely Selfish Sharing.

The main contributions are as follows:
\begin{enumerate}
 \item An examination of the efficacy of previously proposed battery-based sharing algorithms for inter-microgrid sharing, investigated within two distinct scenarios: between counties and within a county settings.
 \item Proposal of a novel sharing algorithm called Selfish Sharing, and assessment of its effectiveness within the aforementioned scenarios. In the proposed selfish sharing algorithm, prosumers share energy if they need it in the future where a percentage of this shared energy would be consumed by the receiving consumer, and the left shared amount would be stored on the consumer's energy storage to be used by the sharing prosumer within a specific amount of time. \textcolor{black}{By using blockchain, the proposed system enables the transactions to be handled between the peers directly without third party involvement, along with matching between the offers, trading requests, and sharing requests in a decentralized manner using smart contracts.}
 \item After analysing the simulation results of the proposed algorithms, we found that energy trading and sharing between microgrids across counties performs better than energy trading and sharing within a county.
 \item Demonstration of the advantageous nature of incorporating energy storage systems to accumulate surplus energy for subsequent discharge during periods of demand, providing tangible benefits to prosumers.
 \item Affirmation of the beneficial impact of both centralized and P2P sharing algorithms in facilitating energy sharing between microgrids.
 \item \textcolor{black}{From a monetary point of view, empirical evidence supports the superiority of the selfish sharing algorithm over P2P and centralized algorithms in the within a county sharing scenario. It also shows a comparative advantage over the centralized sharing algorithm in the between-counties sharing scenario, particularly in terms of decreasing dependency on the grid}.
\end{enumerate} 

In this paper, the term "energy trading" refers to transferring energy for direct monetary return, while "energy sharing" refers to transferring energy in return of a non-monetary form. The paper is organized as follows. Section \ref{rw} provides the literature review and Section \ref{materials_and_methods} demonstrates the materials and methods. Section \ref{experiments} describes the experimental analysis, and Section \ref{discussion} discusses the obtained results. Finally, Section \ref{conc} states the conclusions.

\section{Related Works} \label{rw}

P2P energy trading was investigated in the literature by many researchers.
In \cite{liu2017energy}, researcher presented an innovative model for energy trading among peer-to-peer (P2P) photovoltaic (PV) prosumers within microgrids, incorporating a price-based demand response mechanism. They introduced a dynamic pricing model for energy trading among neighboring PV prosumers to encourage local consumption and enhance economic benefits compared to operating independently. The model's effectiveness was evaluated through a practical case study, demonstrating its capacity to reduce operational costs for PV prosumers by facilitating more economical energy trading compared to individual operation, improving PV energy trading efficiency within the microgrid, enhancing overall energy utilization and sustainability and support demand response by encouraging prosumers to adjust their energy consumption in response to internal price signals, and aligning supply and demand more effectively.

\textcolor{black}{SynergyGrids is a decentralized smart contract-based P2P energy trading system proposed in \cite{aloqaily2022synergygrids} with the aim of reducing the reliance on the utility grid. It implemented a system where energy is traded within and between microgrids using price prediction and matching techniques, where a consortium blockchain facilitated access to peer information and transactions. The trading process was managed by a smart contract that handled user registration, the matching procedure, and resource usage costs.}

Researchers in \cite{zhou2018evaluation} explored P2P energy trading mechanisms to optimize their management on the demand side of power systems by evaluating various P2P energy trading mechanisms within the Great Britain, focusing on residential customers. They introduced a Multiagent-based Simulation Framework for evaluating P2P energy trading mechanisms. This framework is designed to simulate the behaviors of prosumers under various P2P energy trading setups. The framework comprised three types of agents and three corresponding models, designed to mimic the complex interactions within a P2P energy trading system. The agents were Energy Sharing Coordinator Agent (CA) which acts as a market operator within the P2P energy trading region, managing the local energy trades and interacting with the external electricity market/retailer, Prosumer Agents (PAs) which represent individual households or buildings that can produce and consume electricity, and Retailer Agent (RA) which simulates the external electricity market or utility from which the P2P network buys or sells excess energy. The study suggested that, based on the comparisons and index evaluations, the Supply and Demand Ratio (SDR) mechanism emerged as the most promising model for P2P energy trading within the context of Great Britain's residential energy market. It did not only maximize economic benefits for participants but also contributed to a more balanced and self-sufficient energy system. The Mid-Market Rate (MMR) mechanism offered conditional benefits, suggesting its suitability might vary with market conditions, particularly PV penetration levels. Finally, the Bill Sharing (BS) mechanism, while stable, did not significantly improve upon existing paradigms, indicating a need for more innovative approaches to fully tap into the potential of P2P energy trading.

In \cite{cui2019peer}, an efficient energy management strategy within a cluster of smart energy buildings was developed. The research aimed at optimizing energy trading among various types of building such as office, industrial, and commercial. The study proposed utility functions for buildings based on their controllable loads. Then, it introduced a two-stage strategy for energy trading. The first stage aimed to minimize the total social energy cost by optimizing energy trading. The second stage dealt with clearing mutual energy trading through a non-cooperative game approach, where a relaxation-based algorithm is introduced to find the game equilibrium. Their findings underlined the potential of such strategies to enhance energy efficiency, reduce costs, and support the development of sustainable smart grids.

The usage of energy storage to facilitate energy trading is described in \cite{balson2021demand} where researchers introduced Battman. Battman embraces the concept of P2P batteries as a service. Occupants of properties equipped with Battman technology place orders for recharged batteries using the developed application, and individuals from the local general public are compensated for gathering depleted batteries. They are responsible for transporting these batteries to nearby charging points and subsequently delivering the replenished batteries. Prosumers who possess substantial solar power setups can earn income by serving as these charging points within the Battman network. The batteries brought to these designated points are exclusively recharged using solar energy.

Unlike Battman that transfer the energy storage itself to the peers, researchers in \cite{wu2022research} introduced an open and collaborative energy storage cloud platform model. They also proposed a multi-agent shared energy storage transaction model. These innovations were designed to bridge gaps and enable direct transactions between distributed energy sources in the electricity market. In addition, a continuous two-way auction transaction approach was presented, with a tailored quotation strategy for locally distributed energy resources. The model incorporated energy storage state and discharge state, utilizing a continuous auction mechanism for price matching.

Another interesting blockchain-based marketplace for P2P energy trading was introduced in \cite{sahin2023anka}, where the aim of the market was battery-powered devices, such as electric bikes, scooters, and wheelchairs. More specifically, peers with extra energy in their batteries can make energy offers in the market, determining the location, amount of excess energy, the price, and the voltage of their battery. Those information are listed in the decentralized marketplace using a smart contract. Other peers who need to charge their own battery-powered devices, can browse the marketplace using a decentralized application (dApp) and search for nearby offers with compatible batteries and suitable prices.

Facilitating the access to energy to consumers with financial need is an important research topic. In \cite{cali2021novel}, researchers worked on addressing energy poverty problem by introducing energy donation over the blockchain in an P2P energy market. In such market, the energy market had prosumers with PV, and consumers that did not have any PV. A local market aggregator facilitated energy trading and transmission among participants, who used smart devices and a Distributed Ledger Technology (DLT)-based energy trading framework for transactions, dynamic smart contracting, and monetary flow coordination. The utility grid and aggregator engaged in bi-directional energy flow, facilitating trade between the local market and utility. Their work excluded energy storage during successive time periods in the local market, clearing excess energy through collaboration with the utility grid. In this solution, individual donors contributed to the charity, while a social department shares data on recipients in need. In summary, the DLT-based charity system collected and distributed donations to customers with need to energy without being able to afford it.

FederatedGrids proposed a blockchain-based system for P2P energy sharing, where sharing occurs after the trading phase \cite{bouachir2022federatedgrids} to enable consumers without funds to acquire energy through sharing requests. Those sharing requests are then fulfilled by prosumers with surplus energy that wasn't sold during trading. The authors noted the potential future benefits for prosumers but did not implement these benefits in their model. It's crucial to thoroughly discuss and implement these future benefits to prevent exploitation by consumers seeking free energy without offering anything in return.

In \cite{zekiye2024blockchain}, two blockchain-based and battery-enabled trading and sharing algorithms were proposed, namely: Centralized sharing, and P2P sharing, where the sharing is beneficial to clients with no funds. In their work, the clients' batteries enabled the sharing process and eliminated the greedy client problem that could exist in \cite{bouachir2022federatedgrids}. Both algorithms were tested against a dataset of individual houses with data extracted and synthesized from \cite{jhana2019hourly}. 

In this paper, we investigate the usability of the solutions in \cite{zekiye2024blockchain} against a more realistic dataset, which includes energy consumption and production information for clients that are more similar to microgrids rather than individual houses. \textcolor{black}{In addition, we propose a new energy sharing algorithm called selfish sharing, where prosumers prioritize their future energy needs and share energy accordingly. Our proposed battery-based sharing mechanisms, both centralized and P2P, as described in \cite{zekiye2024blockchain} and the selfish sharing in this paper enable consumers with limited funds to access energy while preventing greedy behaviors, which is not addressed in the existing literature \cite{liu2017energy, zhou2018evaluation, cui2019peer}. Although \cite{bouachir2022federatedgrids} discussed energy sharing, it did not present techniques to prevent greediness and motivate prosumers. In our selfish sharing algorithm, consumers must have a partially empty battery managed by the sharing prosumer. Prosumers are incentivized by the ability to reclaim part of the shared energy in the future for their own use. From an energy storage usage perspective, unlike other studies \cite{sahin2023anka, aloqaily2022synergygrids, wu2022research, balson2021demand, cali2021novel}, our system uses energy storage as a tool for energy sharing. Prosumers essentially rent a portion of a consumer's energy storage for a specific time, in exchange for energy the consumer can use immediately.} 


\begin{table}[h]
\caption{Notation used in the equations and algorithms.}
\centering
\begin{adjustbox}{width=1.02\columnwidth,center}
\begin{tabular}{|l|l|}
\hline
\textbf{Symbol} & \textbf{Explanation}           \\ \hline
$p_t$         & P2P trading price at time step t  (EUR) \\ \hline
$up_t$         & Utility grid price at time step t (EUR) \\ \hline
$R_t$             & Number of energy requests in time step t                 \\ \hline
$O_t$         & Number of energy offers in time step t  \\ \hline
$\eta$         & The usable percentage of shared energy  \\ \hline
$\tau$         & The expiry time for the shared energy stored in consumer's energy storage\\ \hline
$m$         & Set of N microgrids\\ \hline
$m_i$         & Microgrid \textit{i}\\ \hline

$m_i.ee$         & Excess energy of microgrid \textit{i}. A negative value indicates a need for energy\\ \hline
$m_i.rc$         & Remaining capacity of microgrid \textit{i}'s energy storage\\ \hline
$m_i.re$         & Reserved energy stored in microgrid \textit{i}'s energy storage\\ \hline
$m_i.balance$         & Balance of microgrid \textit{i}\\ \hline
$m_i.b$ & Battery of microgrid \textit{i}\\ \hline

$sm$ &  The smart contract \\ \hline
$sm.balance$ &  The balance of the smart contract\\ \hline
\end{tabular}
\label{t1}
\end{adjustbox}
\end{table}

\section{Material and Methods} \label{materials_and_methods}

\subsection{Dataset}\label{dataset}
The used dataset is extracted from Enefit's dataset \cite{predictEnergyBehaviorOfProsumers}. The original dataset includes the hourly amount of consumed and produced energies for clients in Estonia between 01, September, 2021 and 29, May, 2023. Clients belong to counties, where 16 counties exist. Clients are classified by the county they belong to, whether they represent a business entity or not, and the product type. Product type represents the contract type with values as Combined, Fixed, General Service, and Spot. Each client represents the aggregation of a number of consumption points by a column called EIC count. Additionally, the capacity of the installed photovoltaic solar panel is provided. As we describe in \ref{sharingalgorithms}, the energy sharing algorithms depend on having an energy storage installed at the client side, thus we randomly generated the energy storage capacity to be relevant to the EIC count. More specifically, each client had the battery capacity randomly generated between 5 and 20 kWh (Kilowatt hours), multiplied by the EIC count, where the range 5 to 20 kW represents the range of battery capacity per house. 

Out of the 69 clients, 57 of them have no missing data within the specified time range, where the other 12 clients have some missing data points. For these 12 clients, missing consumption and production values were set to zero, so they will have no effect on the trading or sharing process.

\begin{figure*}[t]
    \centering
    \includegraphics[width=1\linewidth]{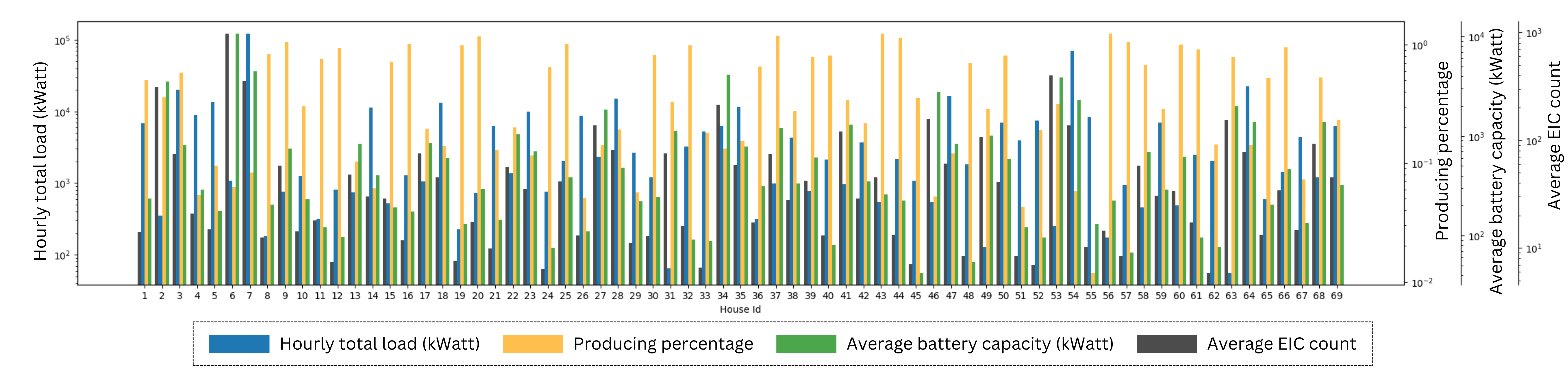}
    \caption{The total load, producing percentage out of total load, average battery capacity, and average EIC count for each grid.}
    \label{fig:per_client}
\end{figure*}

\begin{figure*}[h]
    \centering
    \subfigure[Produced energy per county as a percentage of needed energy]{
        \includegraphics[width=0.32\textwidth]{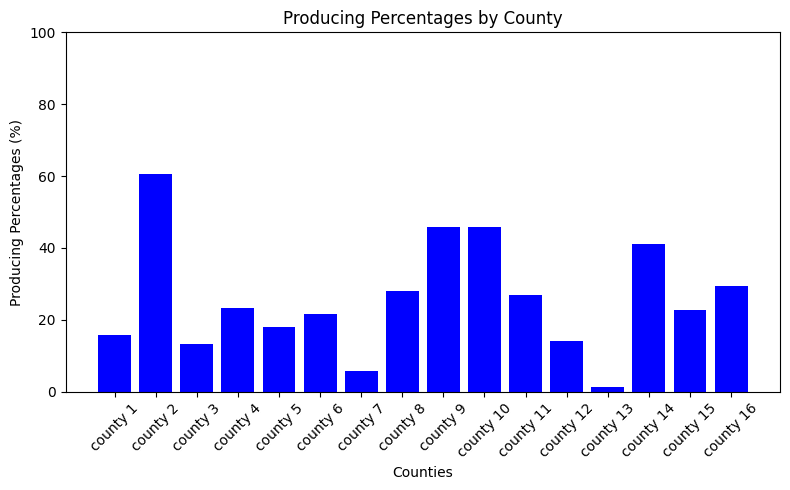}
        \label{fig:producing_percentage_county}
    }
    \hfill
    \subfigure[Clients count per county]{
        \includegraphics[width=0.32\textwidth]{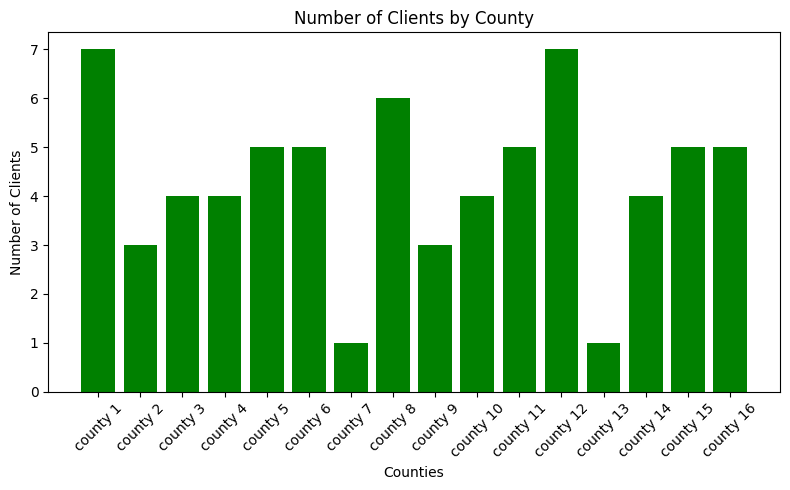}
        \label{fig:number_clients_county}
    }
    \hfill
    \subfigure[The count of business, not business, and the product type distributions]{
        \includegraphics[width=0.3\textwidth]{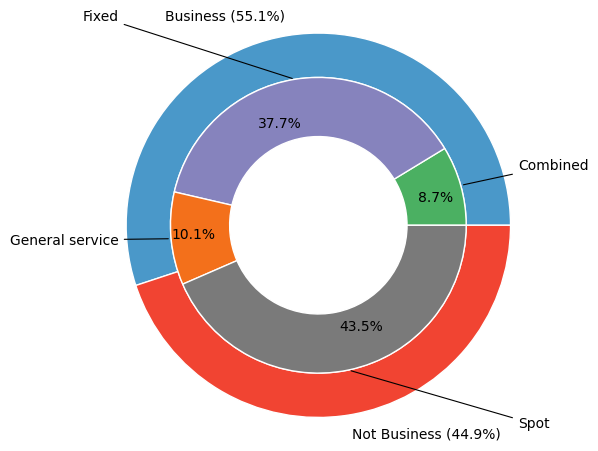}
        \label{fig:ring}
    }
    \caption{Dataset characteristics: produced energy percentage and number of clients per county, product type distributions.}
    \label{fig:both}
\end{figure*}

In our experiments, we considered each client as a microgrid. Figure \ref{fig:per_client} shows the total load, producing percentage out of total load, the average battery capacity, and the average EIC count for each grid we have in the dataset. 
Figures \ref{fig:producing_percentage_county} and \ref{fig:number_clients_county} show the produced energy as a percentage of the needed energy, and the number of clients in each county, respectively. Finally, Figure \ref{fig:ring} shows the percentage of business and non-business clients, along with the distribution of clients per product type.

\subsection{Energy Trading and Sharing}\label{sharingalgorithms}
In this section, we provide a brief overview over our previously proposed centralized sharing and P2P sharing \cite{zekiye2024blockchain}, then we propose a new method called selfish sharing. In this context, consumer will be referring to a microgrid that needs energy, prosumer will be a microgrid that has an excess of energy, and client will be referring to a prosumer or consumer.  

In the centralized, P2P, and selfish sharing algorithms, sharing refers to trading energy for non-monetary returns. More specifically, the clients' energy storage is used as a payment method as follows. The sender, prosumer, will transfer a specific amount of energy to the consumer. The consumer will be storing (1-$\eta$)\% of the received energy in their own energy storage. This stored energy is usable by the sender within a specific amount of time (expiry time) we called $\tau$. After the end of the specified time, any left energy that is stored in the consumer's energy storage and belongs to a sharer will be used by the consumer himself. As a reward for the consumer in return of letting the prosumer store energy in his energy storage, he will consume $\eta$\% of the sent energy directly. For the pricing, we used the same equation that was proposed in \cite{zekiye2024blockchain}, which is shown in \eqref{eq:pricing} where 
\(\text{p}_t\) is the calculated price at the current time step \(t\), \(\text{up}_t\) is the utility price at time step \(t\), \(R_t\), and \(O_t\) are the number of trading requests by consumers and the number of offers made by prosumers at time step \(t\) respectively, and FiT refers to feed-in tariffs, which represents the price that utility grid pays for energy when it is sold to it.

\begin{equation}
\begin{aligned}
\text{p}_t = \max(FiT, \min(\text{up}_t, &  \frac{R_t}{\text{O}_t} \cdot \frac{\text{p}_{t - 1} + \text{p}_{t - 2} + \text{p}_{t - 3}}{3}))
\end{aligned}
\label{eq:pricing}
\end{equation}

\begin{figure}
    \centering
    \includegraphics[width=1\linewidth]{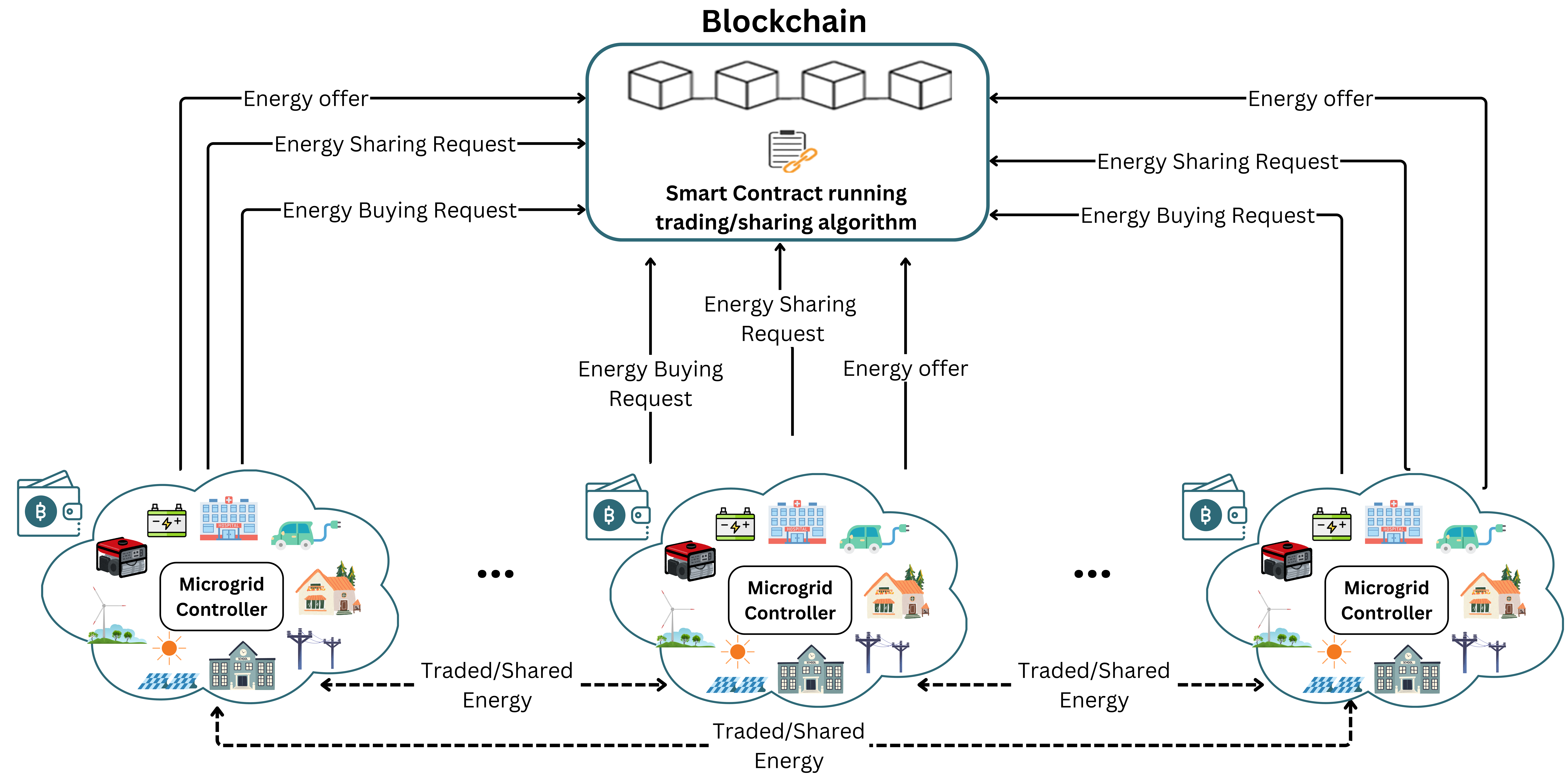}
    \caption{Blockchain-based energy trading and sharing system between microgrids. Dashed lines represent energy flow and solid lines represent data flow.}
    \label{fig:system}
\end{figure}

Figure \ref{fig:system} shows a general representation of our proposed system, where the smart contract manages energy offers, energy buying requests, and energy sharing requests. We can notice from the figure that microgrids are the clients in our system. Each microgrid offers energy, to be shared or traded, when an excess of energy exists, buys energy when it is needed and it has the necessary funds, and asks energy to be shared when energy is needed with lacking funds and having available space in its energy storage. As noticed in the figure, each microgrid consists of distributed energy resources and different loads. More importantly, each microgrid encompasses an energy storage to enable energy sharing, with a capacity relative to its loads.

\subsubsection{Centralized Sharer Entity: C-SE}
In the centralized sharing algorithm, the sharer is a centralized entity that pays to the prosumer directly, share the energy with the consumer, and aims to sell the (1-$\eta$)\ of the shared energy in the next $\tau$ timesteps \cite{zekiye2024blockchain}. 

The algorithm enables trading then sharing as follows. In the trading phase, it matches previously shared energy with energy buying requests, where the profit belongs to the centralized sharer entity. If any energy request is not fully fulfilled, then energy offers get matched with those requests where the centralized sharer gets a fee out of each transaction. Any energy offer that remains unmatched will be allocated to sharing requests, where the sharing phase starts. In this sharing arrangement, a fraction of the shared energy equal to $\eta$ will be consumed by the requesting microgrid, while the remaining fraction (1-$\eta$) will be stored in the requester's energy storage. This stored energy will be available for sale by the centralized sharing entity within a period of $\tau$ timesteps. 

\subsubsection{Peer-to-peer Sharer Entity: P2P-SE}
In the P2P sharing algorithm, the prosumer himself shares the energy directly with the consumer and aims to sell the (1-$\eta$)\% of the shared energy in the next $\tau$ timesteps \cite{zekiye2024blockchain}. 

As in the centralized sharing, this P2P sharing algorithm trades energy then shares it. First, as part of the trading phase, the algorithm matches previously shared energy with energy buying requests, and transfers the profit to the peer entity that shared the energy. If an energy request cannot be completely met, the remaining energy offers are matched with those requests without any fees being charged. Next, the sharing phase starts, where the unmatched energy offers will be allocated to sharing requests. $\eta$\% of the shared energy will be consumed by the requesting microgrid, and the remaining (1-$\eta$)\% will be stored in the requester's energy storage. This stored energy will be saleable by the peer that shared it within $\tau$ timesteps.

\subsubsection{Selfish Sharer Entity: S-SE}
Our proposed method, \emph{Selfish Sharer Entity}, or \emph{selfish sharing}, simulates a selfish prosumer that prefers to share energy to use it in the future for himself, instead of selling it. Unlike the centralized and P2P algorithms, selfish sharing algorithm shares energy then trades it. More specifically, as shown in Algorithm \ref{alg:selfishSharing}, the prosumer will share energy if he needs it in the next $\tau$ timesteps. $\eta$\ of the shared energy will be utilized directly by the consumer who is going to be matched with the prosumer. The left amount, (1-$\eta$)\, will be stored in the matched consumer's battery for $\tau$ timesteps to be used by the same prosumer when needed. Any non-fully-matched offers will be used in the trading phase, where they will be matched with energy buying requests.

In the selfish sharing algorithm, all peers submit either an energy offer, a buying request, or a sharing request according to Algorithm \ref{alg:submit} where the prosumer will use his previously shared energy when needed, before he submits any trading or sharing request.

Any matching between previously shared energies or offers with buying or sharing requests is first-come first-served matching, in the three sharing algorithms: centralized, P2P, and selfish.

\begin{algorithm}[t]
\caption{Matching energy offers with buying and sharing requests in S-SE}
\begin{algorithmic}[1]
\STATE \textbf{input:} \textit{offers, buyRequests, sharingRequests,  $\eta$, $\tau$, $p_t$}

    \FOR{$ offer  \in  offers $}
        \FOR{$ sh  \in  sharingRequests $}
            \STATE $energy = min(offer.amount, sh.amount)$
            \STATE $neededEnergy$ = $offer.prosumer$'s needed energy in the next $\tau$ timesteps $-$ already shared to be reused in the next $\tau$ timesteps
            \IF{$neededEnergy$ $>$ 0}
                \STATE $sharedEnergy = min(energy, neededEnergy)$

                \STATE store $sharedEnergy * (1 - \eta)$ in consumer's battery
                \STATE transfer $sharedEnergy * \eta$ to consumer to use
            \ENDIF
            
        \ENDFOR
    \ENDFOR
\STATE match($buyRequests$, $offers$) 
\end{algorithmic}
\label{alg:selfishSharing}
\end{algorithm}

\begin{algorithm}[]
\caption{Submitting energy offer, buying request, or sharing request}
\begin{algorithmic}[1]
\label{alg:submit}
\STATE \textbf{input:} \textit{h $\leftarrow$ $[h_1,..., h_n]$, $expectedPrice$}
\STATE \textbf{output:} \textit{offers, buyRequests, sharingRequests}

\FOR{$ m_i  \in  h $}
\IF{$ m_i.ee > 0 $ }

\STATE $chargedAmount = min(m_i.ee, m_i.rc)$
\STATE chargeBattery($m_i.b$, $chargedAmount$)

\STATE $m_i.ee = m_i.ee - chargedAmount$
\IF{$ m_i.ee > 0 $ }
\STATE $offers.append($prosumer=$m_i$, $energy\_amount$=$m_i.ee$)
\ENDIF
\ELSIF{$ m_i.ee < 0 $}
\STATE $energyToUseFromSharing = m_i$'s energy shared within previous $\tau$ timesteps

\STATE $chargeUsed = min(m_i.ee, m_i.bc - m_i.re)$
\STATE $m_i.ee = m_i.ee + chargeUsed$
    \IF{$ m_i.ee < 0 $ }
        \IF{$ m_i.balance > m_i.ee * expectedPrice $ }
            \STATE $buyRequests.append($consumer=$m_i$, energy\_amount=$m_i.ee$)
        \ELSIF{$m_i.rc > 0$}
            \STATE $requestedAmount = min(-1 * m_i.ee , m_i.rc)$
            \STATE $shareRequests.append($consumer=$m_i$, requested\_amount=$requestedAmount$ )
        \ENDIF
    \ENDIF
\ENDIF

\ENDFOR
\end{algorithmic}
\label{alg:submit}
\end{algorithm}
\section{Experiments and Results} \label{experiments}
The previously proposed sharing algorithms in \cite{zekiye2024blockchain} along with the selfish sharing algorithm were tested against two different scenarios by making simulations in Python 3. \textcolor{black}{We used Google Colaboratory \cite{google_colab} with approximately 12 GB of RAM and 100 GB of storage capacity. The Pandas library was used for data manipulation, and NumPy \cite{harris2020array} for mathematical operations.
} The matching algorithms in the centralized and P2P sharing algorithms, and the matching process in the selfish Algorithm \ref{alg:selfishSharing} are supposed to run as smart contracts on the blockchain, where they were developed using Solidity. However, for the sake of simulation, we did run all the trading and sharing algorithms off the chain using Python. All three algorithms had expiry time set as 12 timesteps ($\tau$=12), and the usable percentage of shared energy set as 50\%, represented by $\eta$ = 0.5.

The first scenario is \textbf{Between microgrids within a county}, where the trading and sharing takes place between the microgrids in the same county only. On the other hand, the second scenario, \textbf{Between microgrids across counties}, where all microgrids can trade and share energy between each other without any limitation regarding the county. Each client (microgrid) was assigned an initial balance of 10,000 Euro to be used for P2P energy trading only. This balance equals approximately 0.98\% of the average paid money per microgrid to the utility grid when no trading or sharing is applied. The sharing algorithms were compared to two algorithms, trading, and trading with batteries (T\&B). In the trading algorithm, energy is traded between the microgrids where the offers and requests are matched on first-come, first-served manner. In the trading with batteries (T\&B) algorithm, the battery at each microgrid gets charged by the excess energy or discharged if the microgrid needs energy, then the same trading and matching takes place. In the centralized sharing algorithm, the smart contract acquires a fee of 10\% of each transaction to be used in facilitating the haring process.

\subsection{Between microgrids within a county}
The results of simulation for the three sharing algorithms between microgrids within a county can be seen in Table \ref{tab:within_counties_results}, where the average of the results of each county is presented. \textit{Average Energy from Grid} represents the average of total bought energy from the utility grid per county in watt, and the \textit{Average Paid to Grid} is the average of total paid money to the utility grid in Euro. \textit{Average P2P Traded Energy} in watts and \textit{Average Total Earned from P2P Trading} in Euro represents the average amount of P2P traded energy and its price respectively. \textit{Average Energy Wasted/ Sold to Grid} (watt) is the average of the energy that was generated and not used, traded, nor shared. This energy could be sold to the utility grid at a price determined by the utility grid company. \textit{Average Shared by Prosumers} in watt is the average of the total shared energy per county, where $\eta$ of this amount get used by the receiver directly, and (1-$\eta$) is stored in their energy storage to be sold in the future in case of centralized or P2P sharing algorithm, or reused in the case of selfish sharing algorithm, where the entry \textit{Average of Percentage of Sold/Reused Shared Energy} represents the percentage of the shared energy that was resold or reused. The record, \textit{Average Earned from Sharing}, shows the average total of earned money as a result of reselling the shared energy. We can see that the sharing algorithms resulted in decreasing the bought energy from the utility grid, and as a result decreasing the total paid money to it. The reason beyond decreasing the bought energy from the utility grid is utilizing the wasted energy to serve the need of the microgrids. It is noticeable that sharing lead to a decrease in the P2P traded energy as well. To compare between the different algorithms, the difference between the paid money and the earned is calculated, where lower values are better since it refers to less money spent on energy in total. The total earned results from the total sold energy to peers either by trading or sharing, and the total energy sold to grid with a FiT equals the third of the mean price of the simulation result. It is noticeable that having a system with energy storage to store the clean excess energy and use it later is beneficial compared to not having them at all. More than that, although the sharing algorithms have reduced the total bought energy from the grid, it is noticeable that T\&B is more beneficial than them from the monetary point-of-view. It is noticeable that only 51\% of the shared energy to be sold was resold in the centralized sharing algorithm. This percentage was less in the P2P sharing algorithm were it was approximately 26\%. Finally, since we knew when the microgrid is going to need energy in the next $\tau$ timesteps, the percentage of the reused shared energy in the selfish sharing algorithm was 100\%.

\subsection{Between microgrids across counties}
Table \ref{tab:across_counties_results} shows the results of simulation for the three sharing algorithms between microgrids across counties, where N/A means non-applicable. The entries in the table have the same meaning of the entries in Table \ref{tab:within_counties_results} but with the total values of all microgrids, instead of the average of totals between counties. The same observations can be found here, where the sharing algorithms decreased the dependency on the utility grid by utilizing the wasted energy, and as a result decreased the money paid to it. The same effect of decreasing the P2P traded energy was noticed as well. Again, the efficiency of using energy storage is clear compared to the scenarios of not having it. Unlike the \textit{between microgrids within a county} scenario, the P2P sharing algorithm appeared to be better than all other algorithms, where the amount of paid money minus the earned is the least one. Finally, we can noticed that the percentage of resold energy is 94\% and 53\% for the centralized and P2P sharing algorithms, and again 100\% for reused the selfish sharing.

\begin{table*}[t]
\centering

\caption{The average of total results of trading/sharing algorithms between microgrids within a county.}
\begin{tabular}{|l|c|c|c|c|c|c|}
\hline
\textbf{Metric} &\textbf{ No Trading} & \textbf{Trading} & \textbf{T\&B} & \textbf{C-SE }& \textbf{P2P-SE} & \textbf{S-SE}\\
\hline
Average Energy from Grid (watt) & 2.676e+10 & 2.651e+10  &    2.501e+10	 & 2.500e+10 & 2.491e+10  &  2.497e+10 \\ \hline
Average Paid to Grid (EUR) &  4.367e+06 & 4.326e+06  &  4.085e+06  & 4.085e+06 & 4.073e+06 & 4.080e+06\\ \hline

Average P2P Traded Energy (watt) & N/A & 2.553e+08  & 1.494e+08 &  1.455e+08 & 1.362e+08  & 1.483e+08  \\ \hline
Average Earned from P2P Trading (EUR) & N/A & 1.835e+04 	 & 1.158e+04 & 1.122e+04 & 1.032e+04 & 1.143e+04 \\ \hline
Average Energy Wasted/ Sold to Grid (watt) &  3.331e+09 & 3.075e+09  & 2.249e+09  & 2.232e+09  & 2.044e+09 & 2.102e+09 \\ \hline
Average Shared by Prosumers (watt) & N/A & N/A & N/A & 2.087e+07 & 2.185e+08 & 1.039e+08 \\ \hline
Average of Percentage of Sold/Reused Shared Energy & N/A & N/A & N/A & 51\% & 26\% & 100\%  \\ \hline
Average Earned from Sharing (EUR) & N/A & N/A & N/A & 1.496e+03 & 1.761e+03 & N/A \\ \hline
Average of (Total Paid - Total Earned) & 4.1938e+06  & 4.166e+06  &  3.969e+06 & 3.978e+06 &  3.977e+06 &  3.971e+06\\ 
\hline
\end{tabular}
\label{tab:within_counties_results}
\end{table*}

\begin{table*}[t]
\centering

\caption{Results of trading/sharing algorithms between microgrids across counties.}
\begin{tabular}{|l|c|c|c|c|c|c|}
\hline
\textbf{Metric} &\textbf{ No Trading} & \textbf{Trading} & \textbf{T\&B} & \textbf{C-SE }& \textbf{P2P-SE} & \textbf{S-SE}\\
\hline
Total Energy from Grid (watt) & 4.282e+11 & 4.232e+11 &  3.991e+11	 & 3.989e+11 & 3.973e+11 & 3.980e+11 \\ \hline
Total Paid to Grid (EUR)& 6.987e+07 & 6.911e+07 &  6.522e+07 & 6.519e+07 & 6.495e+07 & 6.508e+07 \\ \hline

Total P2P Traded Energy (watt) & N/A & 5.048e+09 & 3.476e+09 & 3.038e+09 & 2.322e+09 & 3.314e+09 \\ \hline
Total Earned from P2P Trading (EUR) & N/A & 3.781e+05	 &  2.658e+05 & 2.421e+05	 & 1.844e+05 & 2.269e+05 \\ \hline
Total Energy Wasted/ Sold to Grid (watt) & 5.3296e+10 & 4.824e+10 & 3.489e+10 & 3.412e+10 & 3.001e+10 & 3.124e+10 \\ \hline
Total Shared by Prosumers (watt) & N/A & N/A & N/A & 1.220e+09 & 6.101e+09 & 5.161e+09  \\ \hline
Percentage of Sold/Reused Shared Energy & N/A & N/A & N/A & 94\% & 53\% & 100\%  \\ \hline
Total Earned from Sharing (EUR) & N/A & N/A & N/A & 5.886e+04 & 1.038e+05 & N/A \\ \hline
Total Paid - Total Earned &  6.710e+07 & 6.660e+07  & 6.340e+07  & 6.342e+07  & 6.339e+07  &  6.346e+07\\ 
\hline
\end{tabular}
\label{tab:across_counties_results}
\end{table*}

\section{Discussion} \label{discussion}
We notice that the usage of energy storage to store the excess generated energy for later reuse is beneficial in both scenarios, between counties, and within a county trading, as it is given in Tables \ref{tab:within_counties_results} and \ref{tab:across_counties_results} by comparing T\&B with the trading and no trading cases. In the \textit{between microgrids within a county} scenario results (Table \ref{tab:within_counties_results}), by comparing the average of total paid minus total earned, the T\&B is the best algorithm, followed by Selfish sharing, P2P sharing, and Centralized sharing. If the aim is to reduce the dependency on the grid, then the P2P sharing is the best, followed by the Selfish sharing, Centralized sharing, and T\&B. On the other hand, in the \textit{between microgrids across counties} scenario results (Table \ref{tab:across_counties_results}), P2P sharing is noted as the best algorithm from the total paid minus earned point-of-view, followed by the Centralized sharing, Selfish sharing, and T\&B. From the total bought energy from the utility grid, the P2P energy is the best, followed by Selfish sharing, Centralized sharing, and T\&B.

In the \textit{between microgrids within a county}, the average of total paid minus earned is reduced by 4.74\% by comparing T\&B with the trading algorithm, and by a decrease of 4.68\% by comparing the Selfish sharing with the trading algorithm. Comparing the decrease in the dependency on the grid, the T\&B decreased it by 5.63\%, and P2P sharing by 6\% when both compared to the trading algorithm. While in the \textit{between microgrids across counties} scenario, the total paid minus earned is reduced by 4.79\% and 4.82\% by comparing the T\&B and the P2P sharing with the trading algorithm respectively. From the dependency on the grid point-of-view, the P2P sharing and T\&B reduced it by 6.11\% and 5.68\%, when compared to the trading algorithm. We can see that the scenario that enables trading and sharing between microgrids across counties is better that the trading and sharing between microgrids within a county. The reason is that when the trading and sharing across counties is enabled, microgrids have more opportunities to trade and share.

\section{Conclusions} \label{conc}
In this paper, we explored the benefits of trading with batteries and two energy sharing algorithms between microgrids in two different scenarios, within a county, and across counties. Additionally, we proposed a new sharing algorithm, namely Selfish Sharing where the energy is shared by prosumers to be reused later by them instead of aiming to reselling it. In both scenarios, trading with batteries was shown to be better than traditional trading both financially and environmentally, by decreasing the dependency on the grid. In the trading and sharing within a county, the trading with batteries was better than other sharing algorithms financially, and the P2P sharing algorithm was better in decreasing the dependency on the grid. In the trading and sharing across microgrids, the P2P sharing algorithm was the best by decreasing the total paid minus earned and decreasing the dependency on the grid as well. In the future, \textcolor{black}{the complexity analysis of the sharing algorithms along with efficiency and feasibility of the smart contracts will be studied. Additionally,} we aim at investigating ways to increase the amount of resold shared energy in both the centralized and P2P sharing algorithms. Finally, we target having a machine learning-based algorithm to forecast the energy need and production of a microgrid within $\tau$ timesteps to be used in the selfish sharing algorithm. 

\section*{Acknowledgment}
This work is supported by TUBITAK (The Scientific and Technical Research Council of Türkiye) 2247-A National Outstanding Researchers Program Award 121C338; and ASPIRE and ZU ViP project EU2105.

\bibliographystyle{IEEEtran}
\bibliography{IEEEabrv,bcca}

\end{document}